\documentclass[aps,
      	       pre,
	             twocolumn,
	             superscriptaddress,
							 showpacs,
							 floatfix
				       ]{revtex4-1}
\usepackage{amsmath, graphicx, enumitem}

\newcommand\psmeetn{P_{\!_-}^{\!^+}}
\newcommand\pnmeets{P_{\!_+}^{\!^-}}

\newcommand\pns{p_{\!_{+\to -}}}
\newcommand\psn{p_{\!_{-\to +}}}

\newcommand\T{\DT_{c}} 
\newcommand\pone{\phi_{c_1}}
\newcommand\ptwo{\phi_{c_2}}

\newcommand\DT{T}

\newcommand\DE{\Delta{}E}
\newcommand\for{\ \forall\ }

\newcommand\Es{E_{\!_{-}}}
\newcommand\En{E_{\!_{+}}}
\newcommand\ns{n_{\!_-}}
\newcommand\nn{n_{\!_+}}
\newcommand\Ns{N_{\!_-}}
\newcommand\Nn{N_{\!_+}}
\newcommand\hs{h_{\!_{-}}}
\newcommand\hn{h_{\!_{+}}}

\newcommand\mus{\mu_{\!_-}}
\newcommand\mun{\mu_{\!_+}}

\newcommand\Ms{M_{\!_-}}
\newcommand\Mn{M_{\!_+}}
\newcommand\Msn{M_{\!_{+-}}}

\newcommand\ms{m_{\!_-}}
\newcommand\mn{m_{\!_+}}
\newcommand\msn{m_{\!_{+-}}}

\newcommand\expns{f_-(t_f)}

\newcommand\Kss{K_{\!_-}^{\!^-}}
\newcommand\Knn{K_{\!_+}^{\!^+}}
\newcommand\Kns{K_{\!_+}^{\!^-}}
\newcommand\Ksn{K_{\!_-}^{\!^+}}
\newcommand\Ks{K_{\!_-}}
\newcommand\Kn{K_{\!_+}}

\newcommand\kns{k_{\!_+}^{\!^-}}
\newcommand\ksn{k_{\!_-}^{\!^+}}
\newcommand\kss{k_{\!_-}^{\!^-}}
\newcommand\knn{k_{\!_+}^{\!^+}}

\newcommand\nns{n_{\!_{+\to -}}}
\newcommand\nsn{n_{\!_{-\to +}}}
\newcommand\deltans{\delta_{\!_{-}}}
\newcommand\deltasn{\delta_{\!_{+}}}

\begin{document}
\title{Macroscopic description of complex adaptive networks co-evolving with
	dynamic node states}

\author{Marc Wiedermann}
\email{marcwie@pik-potsdam.de}
\affiliation{Potsdam Institute for Climate Impact Research --- P.O. Box 60 12
	03, 14412 Potsdam, Germany, EU}
\affiliation{Department of Physics, Humboldt University --- Newtonstr. 15,
	12489 Berlin, Germany, EU}
\author{Jonathan F. Donges}
\affiliation{Potsdam Institute for Climate
	Impact Research --- P.O. Box 60 12 03, 14412 Potsdam, Germany, EU}
\affiliation{Stockholm Resilience Centre, Stockholm University --- Kraftriket
	2B, 114 19 Stockholm, Sweden, EU} 
\author{Jobst Heitzig}
\affiliation{Potsdam Institute for Climate Impact Research --- P.O. Box 60 12
	03, 14412 Potsdam, Germany, EU}
\author{Wolfgang Lucht}
\affiliation{Potsdam Institute for Climate Impact Research --- P.O. Box 60 12
	03, 14412 Potsdam, Germany, EU}
\affiliation{Department of Geography, Humboldt University --- Rudower Chaussee
	16, 12489 Berlin, Germany, EU}
\author{J\"urgen Kurths}
\affiliation{Potsdam Institute for Climate Impact Research --- P.O. Box 60 12
	03, 14412 Potsdam, Germany, EU}
\affiliation{Department of Physics, Humboldt University --- Newtonstr. 15,
	12489 Berlin, Germany, EU}
\affiliation{Institute for Complex Systems and Mathematical Biology,
  University of Aberdeen --- Aberdeen AB24 3FX, UK, EU}
\affiliation{Department of Control Theory, Nizhny Novgorod State University ---
	Gagarin Avenue 23, 606950 Nizhny Novgorod, Russia}
\date{\today}

\begin{abstract}
In many real-world complex systems, the time-evolution of the network's
structure and the dynamic state of its nodes are closely entangled.  Here, we
study opinion formation and imitation on an adaptive complex network which is
dependent on the individual dynamic state of each node and vice versa to model
the co-evolution of renewable resources with the dynamics of harvesting agents
on a social network. The adaptive voter model is coupled to a set of identical
logistic growth models and we mainly find that in such systems, the rate of
interactions between nodes as well as the adaptive rewiring probability are
crucial parameters for controlling 
the sustainability of the system's equilibrium state. We
derive a macroscopic description of the system which provides a general
framework to model and quantify the influence of single node dynamics on the
macroscopic state of the network. The thus obtained framework is applicable to
many fields of study, such as epidemic spreading, opinion formation or
socio-ecological modeling.
\end{abstract}

\pacs{89.75.Fb, 89.75.Hc, 89.65.-s, 87.23.Ge}

\keywords{adaptive networks, opinion formation, co-evolution}

\maketitle

\section{Introduction}
Complex network theory has proven to be a powerful tool for studying
properties, dynamics and evolution of many real-world complex
systems~\cite{Albert2002, Newman2003}.  Of particular interest is to
investigate adaptive or temporal networks and their respective
dynamics~\cite{Gross2008, Gross2009, Holme2012}. Typical processes studied in
this field are epidemic spreading~\cite{Gross2006, May2001,
	Pastor-Satorras2001} or opinion formation, e.g., based on the adaptive voter
model~\cite{Ehrhardt2006, Holme2006}.  Interactions are modeled by randomly
picking a pair of linked nodes and, with fixed probabilities, either changing
the state of one of the two nodes or modifying their neighborhood structure by
adaptive rewiring.  However, recent results have emphasized that opinion
formation and imitation processes in fact do not take place with fixed
probabilities but can depend on the payoff or performance of different
opinion-related choices made by the agents or nodes involved~\cite{Blume1993,
	Szabo1998, Traulsen2010}.

In addition to the structure and dynamics \textit{of} networks there has been a
variety of studies on the dynamics \textit{on} networks, where nodes in the
network represent individual dynamical systems and links indicate directed or
symmetric interactions between them~\cite{Ji2013, Arenas2008}. It has been
suggested that the interplay between the dynamics of and on networks should be
much more investigated, since the dynamics of each of the coupled subsystems is
expected to change significantly when compared to their autonomous
time-evolution~\cite{Gross2008}.  

In this work, we propose a model that combines both aspects. For this purpose
we refine the adaptive voter model so that there is no fixed probability for
pairs of nodes to either imitate each other's  opinion or adaptively rewire
their acquaintance structure. Instead, each node also represents a dynamical
system which, for illustration, is chosen here to be simple and easily
understood if treated in an isolated fashion. In particular, we choose a
logistic growth model, which is a paradigm for the dynamics of a bounded
renewable resource \cite{Perman2003}. Whenever interactions between nodes take
place, the states of the respective dynamical systems are also taken into
account. As a consequence, imitation processes depend explicitly on the nodes'
states as well as on the current network structure. At the same time each of
the nodes' opinions influences a parameter of the local dynamical system.

The proposed model serves as a narrative for possibly emerging dynamics in
co-evolutionary human-nature interactions
\cite{Schellnhuber1998,Schellnhuber1999,lade_regime_2013}. It complements
conceptual studies on the effects of economic growth on the ecospheric
state~\cite{Anderies2013, Kellie-Smith2011} as well as work on resource
exploitation models that take into account the co-evolution of stylized
resource dynamics with a similarly paradigmatic population growth
model~\cite{Brander1998, motesharrei_human_2014}. The proposed model, for the
first time, takes into account individual pairwise interactions of agents on a
social network when studying the stability and dynamics of such intertwined
systems. 

So far, in the context of sustainability science~\cite{schluter_new_2012}, studies on the effect of
different exploitation strategies on the state of a certain ecospheric
component have been carried out by, e.g., studying the extraction of water in
rivers by a network of interconnected harvesters
\cite{Lansing1993,Lansing2009,lansing_perfect_2012}. However, no systematic
analysis of the underlying network structure and resulting dynamics was
performed. In addition, no network dynamics, such as adaptation or imitation
processes, were included in these studies and the focus was mainly set on
studying the state of the ecosphere for different harvesting strategies that
were evolving deterministically in order to optimize all harvesters' payoffs.

In contrast, imitation dynamics with high numbers of agents or players have
been studied in the context of evolutionary game theory
\cite{Szabo1998,Sanfey2007,Traulsen2010,Ebel2002}. However, in no such cases
the dynamics of resources or other externalities has been taken into account
and, hence, no co-evolution of different subsystems has been studied.  Here,
the proposed model serves to illustrate the rich dynamics that may emerge from
the coupling of these different subsystems, even though the complexity in each
of the subcomponents remains manageable.

After the introduction of all key components and processes constituting the
model in Sec.~\ref{sec:model_description} we perform numerical simulations of
the system. In Sec.~\ref{sec:static_network} we first study the case of a
static network and no is adaptation taking place. We find that the system
converges into either a state where all logistic growth models, e.g.,
resources, converge into a state of full depletion or into a state of positive
stock. The latter is to be interpreted as the more sustainable and, hence,
desired outcome of the model. We uncover that the likelihood to converge into
either of the two states is mainly determined by the frequency of interactions
between nodes.

In Sec.~\ref{sec:adaptive_network} we then study the effect of network
adaptation and show that the stability of the system changes in dependence on
the choice of the adaptation frequency. Specifically we deduct that for each
interaction frequency there exists an appropriate rate of network adaptation,
such that the system can be guided into a sustainable state. 

Finally, we derive a low-dimensional set of rate equations for variables that
approximate the model's macroscopic state in
Sec.~\ref{sec:static_network_macroscopics} for the static and in
Sec.~\ref{sec:adaptive_network} for the adaptive case. These equations are
generally applicable to any study of opinion formation or spreading if the
probabilities of changes in node states by imitation are appropriately chosen.
Finally, conclusion are drawn in Sec.~\ref{sec:conclusion}.

\section{Model description}\label{sec:model_description}
Assume a temporal network $G(V, L(t))$ consisting of a fixed set of $N$ nodes
$V=\{v_1, v_2,\ldots , v_N \}$ and an evolving set of links $L(t)$. It is
represented by the time-dependent adjacency matrix $A(t)$. Each node $v_i$
represents a renewable resource stock $s_i(t)$ that obeys a logistic growth
model and is harvested with an effort level $E_i(t)$~\cite{Perman2003},
\begin{align}
	\frac{d}{dt}s_i(t)=a_i s_i(t)(1-s_i(t)/K_i)-q_i s_i(t)
	E_i(t).\label{eqn:log_growth}
\end{align}
For this study, we set the growth rates $a_i = 1$, capacities $K_i = 1$ and
catch-coefficients $q_i = 1$ for all $i=1,\ldots,N$ and measure the time and
stocks in dimensionless quantities. Treating all stocks $s_i$ as evolving under
identical conditions is a strong assumption of the model but allows us to
solely focus on the interplay between network and stock dynamics and its
dependence on a few key parameters.  

The effort is a time-dependent quantity assigned to each node $v_i$ which
defines its current behavioral pattern and changes through imitation of other
nodes. On the one hand, nodes can adopt a \textit{high} effort level $\En >
1$, causing each stock to converge to a stable fixed point $s_+=0$ implying
full depletion of the resource. Alternatively, nodes can choose a \textit{low}
effort level $\Es\in(0,1)$ providing less harvest per unit time initially but
avoiding depletion of the resource stocks since each individual stock $s_i$
then converges to a stable positive fixed point $s_- = 1 - \Es > 0$. The two
possible choices of effort level, $\Es$ (low) and $\En$ (high), are the same
for all nodes and are parameterized by $\DE\in (0, 1)$ such that $\Es = 1 - \DE
$ and $\En = 1+\DE$. At each time $t$ there are $N_{\!_-}(t)$ nodes with
$E_i(t) = \Es$ and $N_{\!_+}(t) = N -N_{\!_-}(t)$ nodes with $E_i(t) = \En$.
The effort then yields for each node $v_i$ an individual harvest $h_i(t) = s_i(t)
E_i(t)$, which constitutes the second term in Eqn.~\eqref{eqn:log_growth}.
From now on we omit the explicit time dependence of the stocks $s_i$, efforts
$E_i$, the adjacency matrix $A$ and the number of low and high effort nodes
$N_{\!_\pm}$, in our notation.

Initially, for each node $v_i$, an individual waiting time $T_i$ is drawn at
random from a Poissonian distribution with density
\begin{align}
p (T_i) = \DT^{-1} \exp(-{T_i}/\DT),
\label{eqn:waiting_time}
\end{align}
which is a typical choice for modeling interaction rates in social
systems~\cite{Haight1967}.  $\DT$ denotes the expected waiting time between two
interactions initiated by the same node $v_i$. Starting from this:
\begin{itemize}[noitemsep]
	\item[(i)] The system as given in Eqn.~\eqref{eqn:log_growth} is
		integrated forward in time for the minimum of all current waiting times
		$T_i$. Then, for the corresponding node $v_i$ (with the smallest $T_i$) a
		neighboring node $v_j$ is drawn uniformly at random.  
	\item[(ii)] If the efforts $E_i$ and $E_j$ of
		$v_i$ and $v_j$ differ:
		\begin{itemize}[noitemsep, leftmargin=*]
			\item[(a)] With a rewiring probability $0\leq \phi \leq 1$, $v_i$ breaks
				its link with $v_j$ such that $A_{ij}=1$ becomes $A_{ij}=0$. Then, a
				new link between $v_i$ and another randomly drawn node $v_k$ with the
				same effort level ($E_i = E_k$) is established such that $A_{ik}=0$
				becomes $A_{ik}=1$. This network adaptation process mimics generally
				observed tendencies to form clusters of individuals with similar
				behavior or social traits. Note that, in contrast to earlier work,
				rewiring only takes place if a randomly drawn neighbor $v_j$ of $v_i$
				shows a different effort, e.g., behavioral pattern~\cite{Holme2006}.  
			\item[(b)] If $v_i$ does not adapt its neighborhood, imitation may happen
				instead (with probability $1-\phi$). The difference in current harvest
				$\Delta h_{ij} = h_j - h_i$ is computed and the node $v_i$
				imitates the current effort level of $v_j$ with a probability given by
				a sigmoidal function $p(E_i \rightarrow E_j) = p(\Delta h_{ij})$ which
			generally is required to be monotonic and continuously differentiable.
			Additionally it must fulfill $p(\Delta h_{ij})\rightarrow 0$ for $\Delta
		h_{ij}\rightarrow -\infty$, $p(\Delta h_{ij}) \rightarrow 1$ for $\Delta
h_{ij}\rightarrow \infty$ and $p(0)=0.5$.  This represents the increasing
likelihood of imitation processes to take place with an increase in the
expected payoff difference~\cite{Traulsen2010}. For our model we set  $p(E_i
\rightarrow E_j) = 0.5(\tanh\Delta h_{ij} + 1)$ which obeys all of the above
requirements.
	  \end{itemize} 
	\item[(iii)] A new waiting time $T_i$ is drawn at random for
		$v_i$ according to Eq.~\eqref{eqn:waiting_time} and step (i) is repeated as
		long as the model has not reached a steady state.
	\item[(iv)] The model reaches (with probability one) a steady state at some
		time $t_f$ when the network divides into one or more components in each of
		which only one choice of effort level is left. 
\end{itemize}
Initially the two possible effort levels are distributed evenly among the nodes
with ratios $\ns(0) = \Ns(0) / N = \nn(0) = \Nn(0) / N = 0.5$. Initial stocks
are set to $s_i(0) = 1$ for all $i=1,\ldots,N$. In the following, we consider
initially Erd\H{o}s-R\'enyi random networks with $N=400$ nodes and a linking
probability of $\rho=\overline k/(N-1)$, where $\overline k=20$ is the average
degree of nodes in the network.

\section{Static network}\label{sec:static_network}
We first study the case of a static network structure with $\phi=0$ (hence,
modeling step (ii)(a) is not implemented at first) and simulate the model
numerically for different combinations of $\DT$ and $\DE$. From this, we derive
a macroscopic approximation of the model constituted from a set of three
coupled differential equations and show its good agreement with the numerical
results.

\subsection{Numerical simulations}\label{sec:static_network_numerics}
\begin{figure}[t]
	\includegraphics[width=.9\linewidth]{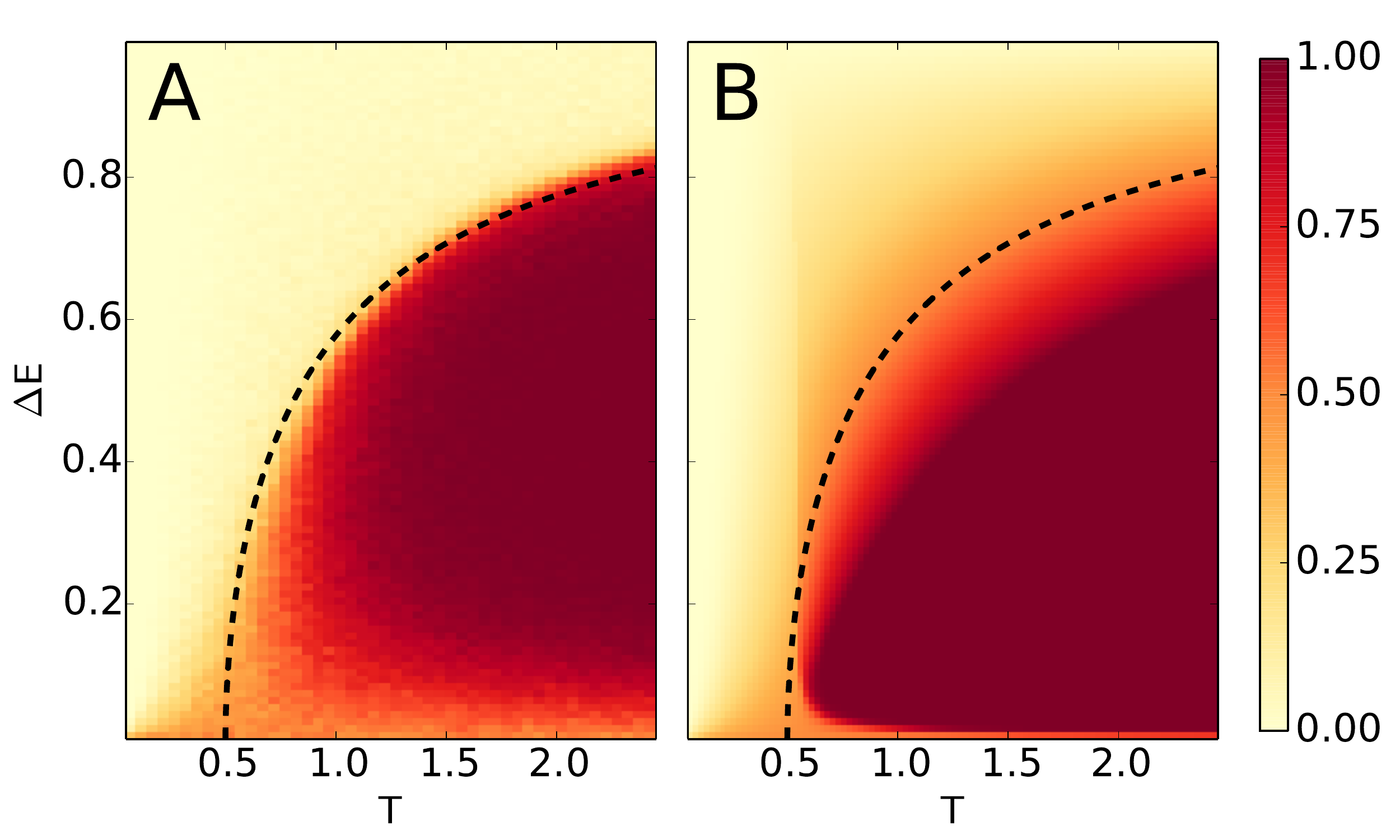}
	\caption{(Color online) (A) The mean fraction $\expns$ of numerical simulations that
		converge to a state where all nodes show a low effort level
		$E_i(t_f)=\Es\for i=1,\ldots,N$ computed over $n=500$ runs for different
		choices of $\DT$ and $\DE$ for a static network with $\phi=0$. (B) The
		value ${\ns}_0$ of the stable fixed point for the fraction $\ns$ of nodes
		with effort level $\Es$ computed from
		Eqs.~\eqref{eqn:05_dns_final}--\eqref{eqn:05_dmun_final}.  The dashed line
		indicates the critical expected waiting time $\T$ which separates the two
		regimes (predominance of nodes using $\En$ (yellow/light) and $\Es$
		(red/dark)).}\label{fig:fig_01}
\end{figure}

Numerical simulations for different combinations of $\DT$ and $\DE$ provide
insights into this system's dynamics. Figure~\ref{fig:fig_01} (A) shows the mean
fraction $\expns$ of model runs that converge to a state where all nodes show a
low effort $E_i(t_f)=\Es\for i=1,\ldots,N$ (using an ensemble of $n=500$
simulations). For small $\DT$ (fast interactions) there is a high probability
for the system to converge to a state where only nodes with a high effort level
$\En$ are present. In this case all resource stocks converge to the stable
fixed point $s_+ = 0$ and become fully depleted. With increasing $\DT$, the
system's expected equilibrium state undergoes a phase transition
in $\expns$. For sufficiently large $\DT$ (slow interactions), the system is
likely to converge to a state where all nodes adopt the effort level $\Es$ and
all stocks converge to a stable fixed point $s_- = 1 - \Es>0$.  This indicates
that the rate of interactions between nodes plays a crucial role in determining
the system's expected equilibrium state. 

The resulting dynamics can be qualitatively understood by considering the
limiting cases of $\DT\rightarrow 0$ and $\DT\rightarrow\infty$. In the first
case, interactions between nodes are expected to happen very fast. Given that
initially all stocks carry the same value $s_i(0) = s_0$ we expect that for
$t\ll 1$ the harvest $\hs$ ($\hn$) of nodes with low (high) effort follows
$\hs(t\ll 1)\propto\Es s_0$ ($\hn(t\ll 1)\propto\En s_0$). This implies that
the difference in harvest between the two different types of nodes is expected
as $\hn(t\ll 1) - \hs(t\ll 1) \propto (\En - \Es) s_0 = 2\DE s_0$. If
interactions happen very fast, the system likely converges into its equilibrium
state at $t_f \ll 1$. Since in this situation we expect $\hn > \hs$, nodes with
low effort are more likely to imitate the high effort rather than the other way
around and, hence, we expect $\expns \rightarrow 0$ for $\DT \rightarrow 0$ (as can
be seen in Fig.~\ref{fig:fig_01} (A)). 

In contrast, for $\DT \rightarrow \infty$ we expect updates between nodes to
happen preferably at times $t\gg 1$. In this case, the stocks of nodes with
high (low) effort can be assumed to have already converged to a fixed point of
$s_+ = 0$ ($s_- = 1 - \En = \DE$) as interactions between nodes start to take
place. Hence, the difference in harvest is expected as $\hs(t\gg 1) - \hn(t\gg
1)=\DE - \DE^2$. Thus, for all $\DE\in (0, 1)$ the harvest of low-effort nodes
exceeds that of nodes with high effort and the system is likely to converge
into a state where all nodes show the low effort and, hence, $\expns
\rightarrow 1$ (red/dark area in Fig.~\ref{fig:fig_01} (A) for high values of $\DT$). 

We note, that $\hs(t\gg 1) - \hn(t\gg 1)=\DE - \DE^2$ varies with $\DE$.
Specifically, in the limiting cases $\DE=0$ and $\DE=1$ we find that the
difference $\hs(t\gg 1) - \hn(t\gg 1)=0$ vanishes and, hence, the system
becomes equally likely to converge into either a state with only low-effort
nodes or only high effort nodes present (see lower right corner and the shift
of the transition point towards higher $\DT$ with increasing $\DE$ in
Fig.~\ref{fig:fig_01} (A)). 

\subsection{Macroscopic approximation}\label{sec:static_network_macroscopics}

Abstracting from pairwise microscopic interactions, we now look at the system
from a macroscopic point of view. Assuming the network to be large and fully
connected at first,
the time evolution of the system's state can be characterized by rate equations
for three quantities: (1) the fraction of nodes $\ns$ with effort level
$\Es$, (2) the mean resource stock $\mus = \langle s_i | E_i = \Es\rangle_i$ of
nodes with effort level $\Es$ and (3) the mean resource stock $\mun =
\langle s_i | E_i = \En\rangle_i$ of nodes with effort level $\En$.  The
fraction of nodes $\nn$ with effort level $\En$ follows from $\nn =
1-\ns$. 

The time evolution of $\ns$ is governed by nodes that change from the low to
the high effort level and vice versa. In particular, in the time interval $(t,
t+dt)$ an infinitesimal fraction of $d\nsn$ ($d\nns$) nodes change their effort
from $\Es$ ($\En$) to $\En$ ($\Es$) which decreases (increases) the fraction of
nodes with low effort $\ns$,
\begin{align}
d\ns = d\nns - d\nsn.
\label{eqn:05_dNs1}
\end{align} 
The interactions between nodes that govern the rates of changes in effort are
driven by the following quantities:
\begin{enumerate}
\item The expected waiting time $\DT$ for a node $v_i$ to interact
	with a randomly drawn neighboring node $v_j$. Correspondingly, 
	the rate of node interactions is taken to be
	$\tau= 1/\DT$.
\item If a node $v_i$ interacts with its neighboring node $v_j$, an imitation
	of effort only takes place if $E_i \neq E_j$. Hence, for a node $v_i$ with
	$E_i = \Es$ ($E_i = \En$) there is to define a probability $\psmeetn$
	($\pnmeets$) that a randomly drawn neighboring node $v_j$ has $E_j = \En$
	($E_j = \Es$). Since a large fully connected network is assumed, this
	probability is given exactly by the current fraction $\nn$ ($\ns$) of nodes
	with high (low) effort $\En$ ($\Es$) and, hence, $\psmeetn = \nn$ ($\pnmeets
	= \ns$).
\item If a node $v_i$ with $E_i=\Es$ ($E_i = \En$) interacts with a neighboring
	node $v_j$ with $E_j = \En$ ($E_j=\Es$), there is a probability $\psn$
	($\pns$) that $v_i$ takes up the effort level $E_j$ of $v_j$.  This
	probability is governed by the difference in harvest $\Delta h_{ij}$ between
	$v_j$ and $v_i$. For the macroscopic description, the individual pairwise
	interactions are replaced by aggregated quantities. Therefore $\psn$ ($\pns$)
	is computed as the \textit{expected} probability for a node $v_i$ with low
	(high) effort to adopt the high (low) effort given that it interacts with a
	node $v_j$ that currently has $E_j = \En$ ($E_j = \Es$). This quantity is
	then dependent on the expected stocks at nodes with low and high 
	effort, which is derived below in detail.
\end{enumerate}
This yields $d\nsn$ and $d\nns$ as the
product of all three factors introduced above,
\begin{align}
d\nsn &= \ns \tau \nn \psn dt \\
d\nns &= \nn \tau \ns \pns dt \\
\Rightarrow \frac{d\ns}{dt} &= \tau\ns\nn(\pns - \psn).\label{eqn:dns_final}
\end{align}
The two quantities still remaining to be evaluated, are the expected
probabilities $\pns$ ($\psn$) for nodes with a high (low) effort level to change
to the opposite level. It is obtained as the expected probability for nodes in
the network to take up its neighbor's effort, 
\begin{align}
\pns&=\langle P(E_j \rightarrow E_k)\ |\ E_j=\En,\ E_k=\Es
\rangle_{j,k}\nonumber \\
    &=0.5\langle\tanh(\Delta h_{jk}\ |\ E_j=\En,\ E_k=\Es)\rangle_{j,k}+0.5\nonumber\\
&\cong0.5\langle\Delta h_{jk}\ |\ E_j=\En,\ E_k=\Es\rangle_{j,k}+0.5\nonumber\\
    &=0.5(\Es\langle s_k\ |\ E_k=\Es\rangle_k-\En\langle s_j\ |\
		E_j=\En\rangle_{j})\nonumber\\&\quad+0.5\nonumber\\
    &=0.5(\Es\mus-\En\mun)+0.5\label{eqn:static_pns}\\
\psn&=0.5(\En\mun-\Es\mus)+0.5\label{eqn:static_psn}.
\end{align}
Here, we performed a linear expansion of the hyperbolic tangent, $\tanh
x=x+O(x^3)$, assuming that differences in harvest remain small. 

The time evolution of either of the two average stocks $\mus$ and $\mun$ is
governed by two terms. First, each individual stock $s_i$ follows the logistic
growth model and so do the average quantities. Second, the value of each of the
two average stocks changes according to the fact that the nodes modify their
effort from $\Es$ to $\En$ and vice versa during the time interval $(t, t+dt)$.
This yields
\begin{align}
d\mus&=d\langle s_k\ |\ E_k=\Es\rangle_k\nonumber\\
     &=\langle ds_k\ |\ E_k=\Es\rangle_k\nonumber\\
     &=dt\ \langle s_k(1-s_k)-E_ks_k\ |\ E_k=\Es\rangle_k+\deltans\nonumber\\
		 &=dt\mus-dt\langle s_k^2 \ |\ E_k=\Es\rangle_k
		 -dt\Es\mus+\deltans\nonumber\\
     &=dt(\mus(1-\mus-\Es)-\mus^{(2)})+\deltans\label{eqn:stock1}\\
d\mun&=dt(\mun(1-\mun-\En)-\mun^{(2)})+\deltasn\label{eqn:stock2}.
\end{align}
Here $\mus^{(2)}$ and $\mun^{(2)}$ denote the variances in the two types of
stocks. $\deltans$ ($\deltasn$) indicate the net change in the average stock
as nodes with high (low) effort change their effort to the opposite choice
during $(t,t+dt)$. The fraction of nodes $d\nns$ ($d\nsn$) that
change their effort from $\En$ to $\Es$ ($\Es$ to $\En$) during $(t,t+dt)$ is
assumed to be small compared to the fraction of nodes which
already hold the low (high) effort, $d\nns\ll\ns$ ($d\nsn\ll\nn$). Hence, the
respective contribution to the dynamics of $\mus$ ($\mun$) as nodes change
their effort is also assumed to be small, $d\nns\mun\ll\ns\mus$
($d\nsn\mus\ll\nn\mun$). This allows for a first-order expansion of the stock's
time evolution, such that 
\begin{widetext}
\begin{align}
	\mus+\deltans&=\frac{(\ns - d\nsn)\mus+d\nns\mun}{\ns-d\nsn+d\nns}\nonumber\\
	&
	\cong\left.\frac{(\ns-d\nsn)\mus+d\nns\mun}{\ns-d\nsn+d\nns}\right|_{(d\nsn,d\nns)=(0,0)}\nonumber\\
	&\quad +\left.\frac{-\mus(\ns-d\nsn+d\nns)+((\ns-d\nsn)\mus+d\nns\mun)}{(\ns-d\nsn+d\nns)^2}
	\right|_{(d\nsn,d\nns)=(0,0)}d\nsn\nonumber\\
	&\quad +\left.\frac{\mun(\ns-d\nsn+d\nns)-((\ns-d\nsn)\mus+d\nns\mun)}{(\ns-d\nsn+d\nns)^2}
	\right|_{(d\nsn,d\nns)=(0,0)}d\nns\nonumber\\
	&=\frac{\ns\mus}{\ns}+\frac{-\mus\ns+\ns\mus}{\ns^2}d\nsn+\frac{\mun\ns-\ns\mus}{\ns^2}d\nns
	=\mus+\frac{\mun-\mus}{\ns}d\nns\\ \Rightarrow
	\deltans&=(\mun-\mus)\nn\tau\pns dt\label{eqn:deltans}\\ 
	\deltasn&=(\mus-\mun)\ns\tau\psn dt.\label{eqn:deltasn}
\end{align} 
\end{widetext}
Putting this back into~\eqref{eqn:stock1} and~\eqref{eqn:stock2}
yields 
\begin{align} d\mus&=dt
	(\mus(1-\mus-\Es)-\mus^{(2)})\nonumber\\&\quad+dt(\mun-\mus)\nn\tau\pns\\ d\mun&=dt
	(\mun(1-\mun-\En)-\mun^{(2)})\nonumber\\&\quad+dt(\mus-\mun)\ns\tau\psn.  
\end{align} 
In the scope of this work, in to order to close the set of equations that
describe the systems dynamics, we assume the respective variances $\mus^{(2)}$
and $\mun^{(2)}$ to vanish. Taking into account higher moments in the dynamics
of the stocks and investigate its influence on the resulting fixed points
remains as a task for future research. 

In summary, we find a set of three coupled ordinary differential equations that
define the time evolution of the static network model:
\begin{align}
\frac{d\ns}{dt} &=  \tau \nn \ns (\pns - \psn)\label{eqn:05_dns_final}\\
\frac{d\mus}{dt} &= \mus(1 - \mus - \Es) + \tau (\mun - \mus)  \nn \pns
	\label{eqn:05_dmus_final} \\
\frac{d\mun}{dt} &= \mun(1 - \mun - \En) + \tau (\mus - \mun)  \ns \psn
	\label{eqn:05_dmun_final}.
\end{align}

\subsection{Fixed points and stability}\label{sec:stability}
\begin{figure}[t]  
    \begin{center}
    \includegraphics[width=0.9\linewidth]{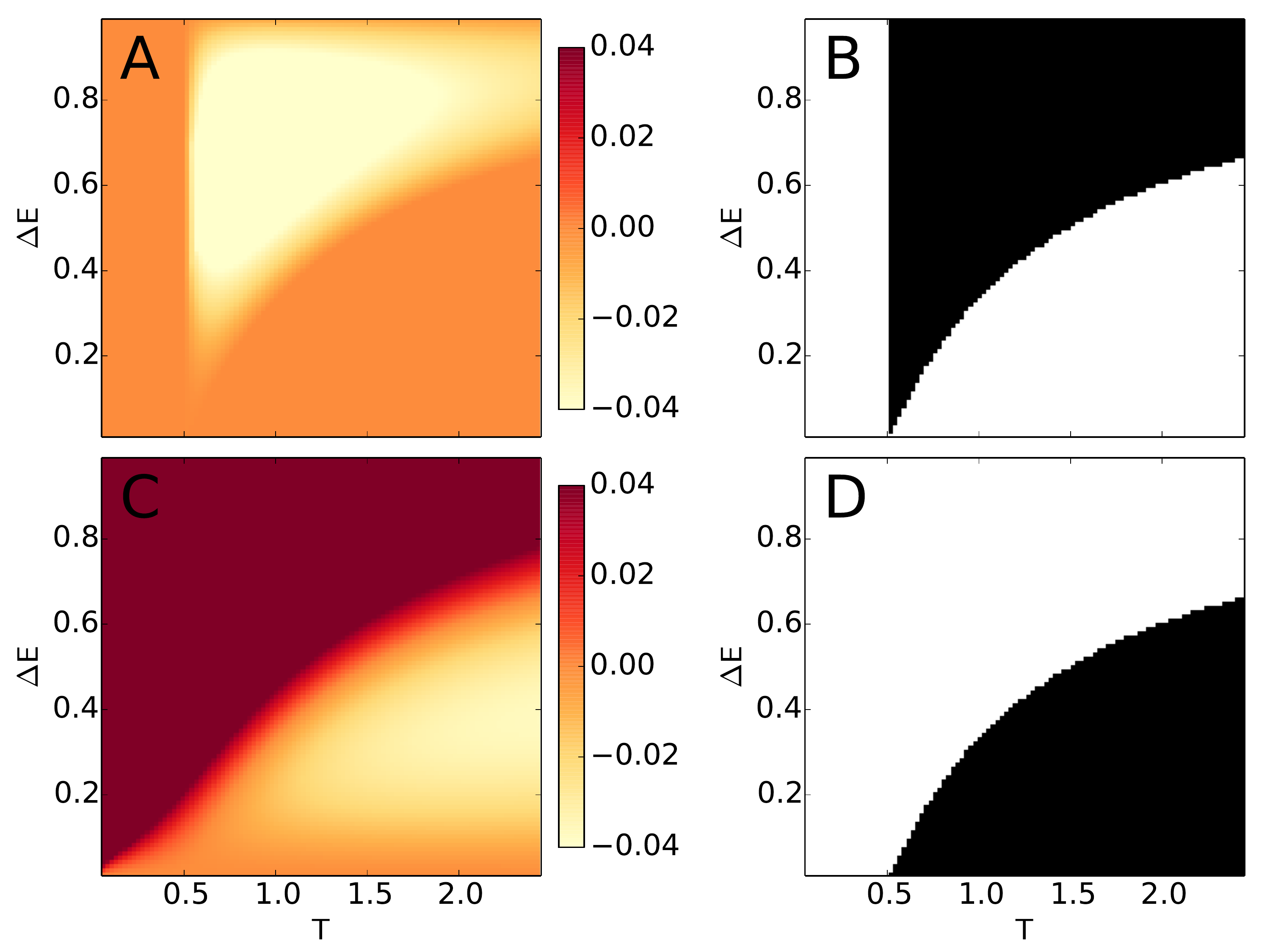}
		\caption{(Color online) The largest eigenvalue $\lambda_1$ for the two fixed points $P_3$
			(A) and $P_4$ (C) (see also Eqs.~\eqref{eqn:fp3} and~\eqref{eqn:fp4})
			depending on $\DE$ and $\DT$. The black area in (B) indicates the domain
			in parameter space for which $\lambda_1$ computed for $P_3$ is negative
			and, hence, $P_3$ is stable. (D) shows the same properties for $P_4$. The
			regimes for which either of the two fixed points is stable are
			complementary. Further it should be noted that for $\DT<0.5$ neither of the
			two fixed points is stable, but center manifold as given in
			Eqn.~\eqref{eqn:05_center_manifold} exists in this
			regime.}\label{fig:fig_02}
     \end{center}
\end{figure}

We obtain all fixed points $P_i=({\ns}_0,{\mus}_0,{\mun}_0)$ of the dynamical
system given in Eqs.~\eqref{eqn:05_dns_final}--\eqref{eqn:05_dmun_final} as:
\begin{widetext}
\begin{align}
P_1&=\left({\ns}_0=0,\ {\mus}_0=\frac{1-\Es-0.5\tau}{1+0.5\tau\Es},\ {\mun}_0=0\right)\\
P_2&=\left({\ns}_0=1,\ {\mus}_0=0,\ {\mun}_0=\frac{1-\En-0.5\tau}{1+0.5\tau\En}\right)\\
P_3&=\left({\ns}_0=\frac{2(\Es\frac{1-0.5\tau}{\Es+\En}+\En-1)}{\tau(\frac{\En}{\Es}-1)},\
	{\mus}_0=\En\frac{1-0.5\tau}{\Es+\En},\
	{\mun}_0=\Es\frac{1-0.5\tau}{\Es+\En}\right)\label{eqn:fp3}\\
P_{4}&=\left({\ns}_0=1,\ {\mus}_0=1-\Es,\
	{\mun}_0=\frac{-b}{2a}+\sqrt{\left(\frac{b}{2a}\right)^2+\frac{c}{a}}\right)\label{eqn:fp4}\\
P_{5}&=\left({\ns}_0=1,\ {\mus}_0=1-\Es,\
	{\mun}_0=\frac{-b}{2a}-\sqrt{\left(\frac{b}{2a}\right)^2+\frac{c}{a}}\right)\\
a&=0.5(-2-\En\tau)\nonumber\\
b&=1-\En+0.5\tau((1-\Es)\En+\Es-\Es^2-1)\nonumber\\
c&=0.5\tau(1-\Es)(\Es-\Es^2-1).\nonumber
\end{align}
\end{widetext}
In addition, there exists a manifold which also satisfies $\frac{d\ns}{dt} =
\frac{d\mus}{dt} = \frac{d\mun}{dt} = 0$ and is defined by
\begin{align}
P_\alpha = ({\ns}_0=\alpha,\ {\mus}_0=0,\ {\mun}_0=0),\ \alpha\in\lbrack
0,1\rbrack\ \label{eqn:manifold}
\end{align}
For all fixed points given above we compute the largest eigenvalue $\lambda_1$ of the
corresponding Jacobian matrix evaluated at the respective point. Only the two fixed
points $P_3$ and $P_4$ have a negative largest eigenvalue $\lambda_1 < 0$ and,
hence, are stable for choices of parameters $\DE$ and $\DT>0.5$ (note that
again: $\Es=1-\DE$, $\En=1+\DE$ and $\tau=1/\DT$) (Fig.~\ref{fig:fig_02}).

To investigate the system's dynamics in the regime $\DT < 0.5$, the stability
on the 1-dimensional manifold defined by all points that fulfill Eq.~\eqref{eqn:manifold} is assessed. Analytically computing the three eigenvalues of
the Jacobian matrix on the manifold as a function of the parameter $\alpha$ yields
\begin{widetext}
\begin{align}
\lambda_0 &= 0\\
\lambda_\pm(\alpha)&=1-\frac{\En+\Es}{2}-\frac{\tau}{4}\pm \frac{1}{2}
\sqrt{2\alpha\En\tau-2\alpha\Es\tau+\En^2-2\En\Es-\En\tau+\Es^2+\Es\tau+\frac{\tau^2}{4}}.\label{eqn:cond_alpha}
\end{align}
\end{widetext}
A first observation is that 
$\lambda_+(\alpha)\geq\lambda_-(\alpha)$ holds. Since $\lambda_0=0$, it
is obvious that not all eigenvalues can be negative. However, if $\lambda_0=0$
is the largest eigenvalue of the system, all choices of $\alpha$ for which
$\lambda_+(\alpha) \leq \lambda_0$ define a center manifold,
\begin{align}
\lambda_+(\alpha)\leq 0\ \text{if}\ \alpha &\leq
\frac{1}{2}-\DT\DE\label{eqn:05_alpha}
\end{align}
Hence,
\begin{align}
\nu(\alpha) &= (n_{s0} = \alpha, \mu_{s0}=0, \mu_{n0} = 0)\label{eqn:05_center_manifold}\\
\alpha & \in\left [0, \frac{1}{2}-\DT\DE \right ]\nonumber
\end{align}
defines a center manifold where the system's stability cannot be assessed by
linear stability analysis. A detailed study of the system's stability in this
regime is beyond the scope of this work and not necessarily needed to
understand the general dynamics of the macroscopic description proposed here.
Numerically integrating the system for choices of parameters taken from the
center manifold, however, reveals good agreement between the microscopic and
macroscopic model representation (Fig.~\ref{fig:fig_01}).
An investigation by means of a higher-order stability analysis might, however,
yield further insights into the processes that cause both resource stocks
${\mus}_0={\mun}_0=0$ to be fully depleted in the regime of the center manifold.

In conclusion, we note that for each choice of parameters only one of the fixed
points $P_3$ and $P_4$ can be the unique stable fixed point of the system
(Fig.~\ref{fig:fig_02}).
Figure~\ref{fig:fig_01} (B) displays the value of the stable fixed point's
${\ns}_0$-component as a function of $\DT$ and $\DE$. The results are in good
agreement with the numerical findings (Fig.~\ref{fig:fig_01} (A)). Due to the
first-order approximation, the transition from a predominance of nodes with
$\En$ to nodes with $\Es$ with increasing $\DT$ is not as sharp as for the
numerical simulations.  However, a good estimate for the critical value $\T$ of
$\DT$ at which the transition takes place can be found by setting ${\ns}_0(\T)
= 0.5$ in Eq.~\eqref{eqn:fp3} which yields $\T(\DE) = \frac{1 + \DE^2
}{2-2\DE^2}$ (dashed line in Fig.~\ref{fig:fig_01}).

\section{Adaptive network}\label{sec:adaptive_network}
In the following, we consider additionally network adaptation processes with $\phi>0$ (hence,
modeling steps (ii)(a) and (b) both take place with a relative frequency depending on
the rewiring probability $\phi$).  For all results presented from here on, the
two available choices of effort levels are fixed by setting $\DE=0.5$.
\subsection{Numerical simulations}
Numerical simulations with the same initial conditions as in the static case
for different combinations of $\phi$ and $\DT$ reveal a division of the
parameter space into regimes of different expected outcomes as the model
reaches its steady state (Fig.~\ref{fig:fig_03} (A)). As for $\phi=0$, fast
interactions (i.e., low values of $\DT$) lead to a large fraction of nodes
carrying $\En$. The transition between the two behavioral
patterns with increasing $\DT$ remains sharp.  However, depending on the choice
of $\phi$, the value of the critical waiting time $\T$, at which the system
transfers from a state with a predominance of nodes with low effort to a state
with a predominance of nodes with high effort, decreases with increasing
$\phi$.  Conversely, this implies that for all $\DT \gtrsim 0.3$ there is an
appropriate choice of $\phi \in [\pone, \ptwo]$ so that all nodes are likely to
adopt the effort level $\Es$. In the limiting case of $\phi=1$ the
expected fraction of nodes with $\Es$ equals the initial
fraction $\ns(0) = 0.5$ for all choices of $\DT$ due to the network's
fragmentation into components of nodes sharing the same effort. 

\subsection{Macroscopic approximation}
The macroscopic approximations
\eqref{eqn:05_dns_final}--\eqref{eqn:05_dmun_final} can be extended to also
include the effects of network rewiring. For this, we introduce two additional
variables describing the macroscopic state of the network. 
The time evolution of the fraction of nodes $\ns$ with low
effort is recalled (analogously to Eq.~\eqref{eqn:dns_final}) as 
\begin{align}
\frac{d\ns}{dt} = \tau(\nn\pnmeets\pns - \ns\psmeetn\psn).
\label{eqn:dns_adaptive_general}
\end{align}
Given that a node $v_i$ initializes an interaction and the randomly drawn
neighboring node $v_j$ employs a different effort, $E_i\neq E_j$, there exists
the adaptive rewiring probability $\phi\in[0,1]$ for $v_i$ to break its
connection with $v_j$ and establish a link with another randomly drawn
node $v_k$ in the network that is employing the same effort as $v_i$
($E_k=E_i$) and is not yet connected to node $v_i$. With probability $1-\phi$,
imitation of efforts takes place which has already been implemented in the macroscopic
description of the static network.  To account for the adaptive
rewiring process, the interaction rate $\tau$ needs to be refined such that it
no longer represents the rate of node interactions alone, but the rate of
interactions which lead to imitation,
\begin{align}
\tau = \frac{1-\phi}{\DT}.\label{eqn:tau_adaptive}
\end{align}
Likewise the ratio $\rho$ of all node interactions that lead to adaptive
rewiring needs to be defined. Since each node is expected to interact at a rate
$1/\DT$ it follows that
\begin{align}
\rho = \frac{\phi}{\DT}.
\end{align}
For adaptive rewiring to take place, the network cannot be fully connected.
Therefore, the previous definitions of $\pnmeets=\ns$ and $\psmeetn=\nn$ for
two nodes of different effort to interact no longer hold for the derivations to
be performed here. 
\begin{figure}[t]
\includegraphics[width=.9\linewidth]{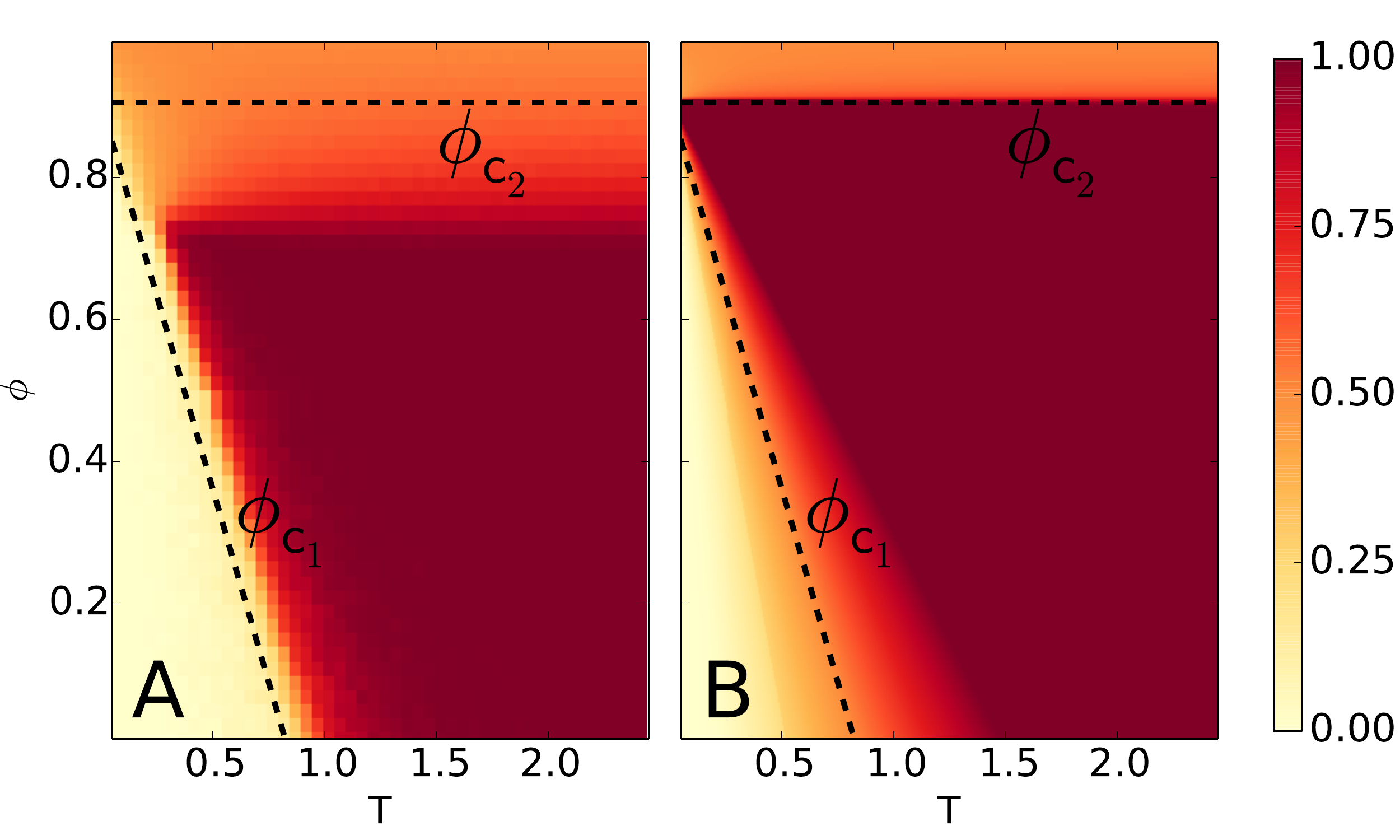}
\caption{(Color online) (A) Mean fraction
		of nodes $\expns$ with effort level $\Es=1-\DE=0.5$ for different choices
		of $\DT$ and $\phi$ obtained from an ensemble of $n=500$ numerical simulations as the
		system reaches its steady state. (B) Value of the stable fixed point for
		the fraction of nodes with effort level $\Es$ computed from the set of differential
		equations~\eqref{adaptive_1}--\eqref{adaptive_2}. }\label{fig:fig_03}
\end{figure}

The total number of $M$ links in the network splits into $\Ms$
($\Mn$) links connecting two nodes with low (high)
effort and $\Msn$ links connecting two nodes of different efforts, such that
\begin{align}
M &= \frac{N\overline k}{2} = \Ms+\Mn+\Msn\\
\Rightarrow \frac{dM}{dt} &= \frac{d\Ms}{dt} + \frac{d\Mn}{dt} + \frac{d\Msn}{dt} = 0.
\end{align}
Additionally
let 
\begin{align}
	\Kss = \frac{2\Ms}{\Ns}\label{eqn:Kss}
\end{align}
denote for nodes with low effort the average number of neighbors
with the same effort. Likewise,
\begin{align}
	\Ksn &= \frac{\Msn}{\Ns}
\end{align}
represents for nodes with low effort the average number of neighbors with high
effort. These two quantities constitute the average degree of nodes with low effort as 
\begin{align}
	\Ks=\Kss+\Ksn=\frac{\Msn+2\Ms}{\Ns}.
\end{align}
Likewise the average degree $\Kn$ of nodes with high effort is obtained from
\begin{align}
	\Knn &= \frac{2\Mn}{\Nn}\label{eqn:Knn}\\
	\Kns &= \frac{\Msn}{\Nn}\label{eqn:Kns}\\
	\Kn&=\frac{\Msn+2\Mn}{\Nn}.
\end{align}
For a node $v_i$ currently having a low effort $E_i=\Es$ the probability
$\psmeetn(v_i)$ to draw a neighbor $v_j$ with different effort at random is given
as
\begin{align}
	\psmeetn(v_i) = \frac{\ksn(v_i)}{k(v_i)}.
\end{align}
$\ksn(v_i)$ is the number of neighbors of node $v_i$ that employ the high
effort. $k(v_i)$ denotes the degree of node $v_i$. Since for the macroscopic
description the pairwise microscopic interactions between nodes are
approximated by the average dynamics, we compute the average probability
$\psmeetn$ for a node $v_i$ with low effort to interact with a node employing
the high effort. Since the network is initialized as an Erd\H{o}s-R\'enyi
random network and it is further equally likely for all nodes 
with the same effort to connect to or disconnect from other nodes by
random rewiring, we perform a heterogeneous mean-field approximation and assume the average degree $k(v_i)$ to be the same for all
nodes with low effort, $k(v_i) = \Ks\for i \in \{1,\ldots,N\ |\
E_i=\Es\}$~\cite{vespignani_modelling_2012,castellano_thresholds_2010}.
Thus
\begin{align}
\psmeetn&=\langle\psmeetn(v_i)\ |\ E_i = \Es
\rangle_i\nonumber =\left\langle\frac{\ksn(v_i)}{k(v_i)}\ \middle|\ E_i=\Es\right\rangle_i\nonumber\\
&=\left\langle\frac{\ksn(v_i)}{\Ks}\ \middle|\ E_i=\Es\right\rangle_i
			   =\frac{\Ksn}{\Ks}\nonumber\\
			   &=\frac{\Msn}{2\Ms+\Msn}. 
\end{align}
Instead of the actual number of $M$ links in the network we define the
corresponding per node link density
\begin{align}
m&=\frac{M}{N}=\frac{\Msn}{N}+\frac{\Ms}{N}+\frac{\Mn}{N}\nonumber\\&=\frac{\overline k}{2}=\msn+\ms+\mn
\end{align}
which is independent of the number of nodes $N$ in the network. The probability for a node with
low (high) effort to interact with a node of high (low) effort is then given by
\begin{align}
\psmeetn=\frac{\msn}{2\ms+\msn}\label{eqn:adaptive_psmeetn}\\
\pnmeets=\frac{\msn}{2\mn+\msn}\label{eqn:adaptive_pnmeets}
\end{align}
and is fully determined by the per node densities of links $\msn$, $\mn$ and
$\ms$. 

Generally, the time evolution of the total number of links between nodes of low
effort is governed by imitation and adaptation. First, we focus on the process
of adaptation. Since links between nodes of the same effort can only be
established but not removed via the process of adaptation, the contribution of
this process to the total number of links between low-effort nodes $\Ms$ only
causes it to increase. This positive contribution is
\begin{align}
	\frac{d\Ms}{dt} \sim \rho \Ns\psmeetn \label{eqn:adaptation} 
\end{align} and
is explained as follows: In each time interval $(t, t+dt)$ there is a total
number of $\Ns$ nodes, which with probability $\rho$ initiate an interaction
that leads to adaptive rewiring. Adaptive rewiring then takes place if a
randomly drawn neighbor $v_j$ of the considered node $v_i$ employs the high
effort. As defined above, this happens with probability $\psmeetn$.

\begin{figure}[t]
    \begin{center}
    \includegraphics[width=0.9\linewidth]{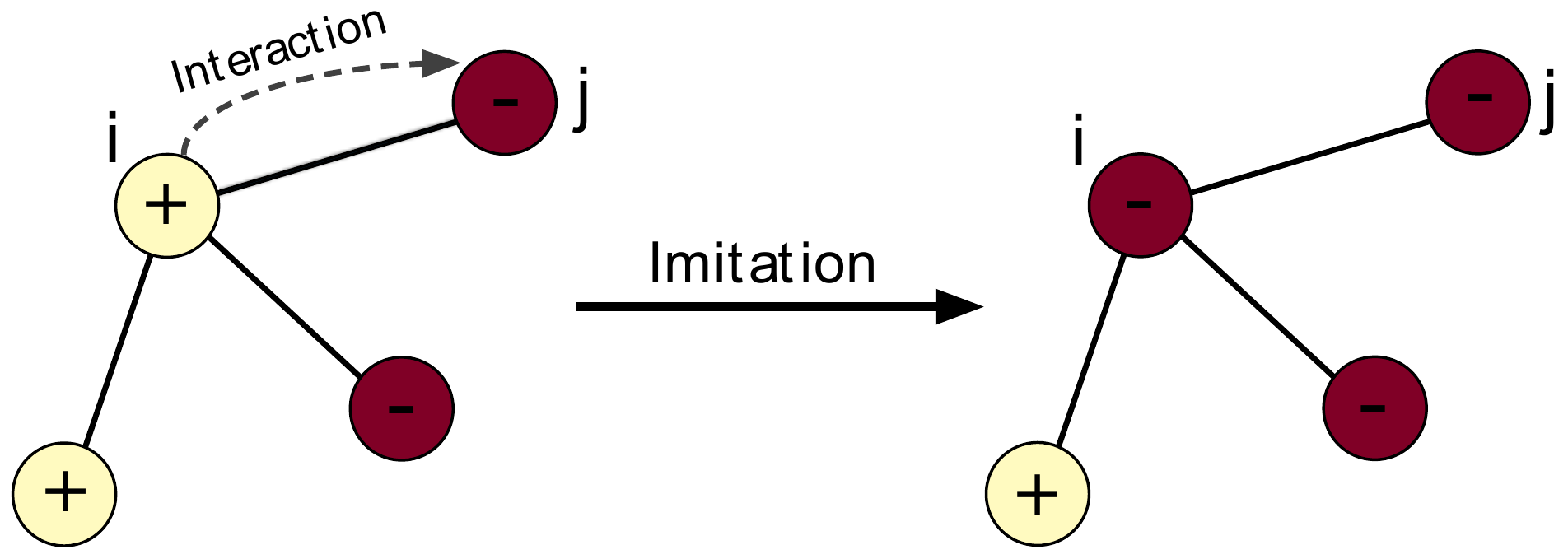}
    \caption{(Color online) Illustration of the influence of the imitation of effort on the
			different numbers of link types in the network. A node $v_i$ with the
			high effort $E_i=\En$ (indicated in orange) interacts with a node $v_j$
			with low effort $E_j=\Es$ (red/dark). Node $v_i$ may then imitate the effort
			of node $v_j$, $E_i\rightarrow \Es$. The number of links between nodes
		with low (high) effort $\Ms$ ($\Mn$) then increases (decreases) by the number
		$\kns(v_i)$ ($\knn(v_i)$) of neighbors of
		$v_i$ that show the low (high) effort.}\label{fig:fig_04}
    \end{center}
\end{figure}
The second contribution to the time evolution of $\Ms$ is given by imitation, which takes place
at rate $\tau$. Generally, there is one term causing an increase in links
between nodes with low effort and one term causing its decrease. First, assume a node $v_j$ with $E_j=\En$ to imitate the
low effort $\Es$ from one of its neighboring nodes $v_i$ with $E_i=\Es$. The number of links between nodes of low effort then increases by the number $\kns(v_j)$ of all neighbors of node $v_j$ that employ the low
effort (Fig.~\ref{fig:fig_04}). Again, by performing a heterogeneous mean-field
approximation and assuming the number of neighbors for
individual nodes to be represented by the respective average number of
neighbors, we set
\begin{align}
	\kns(v_j)=\Kns=\frac{\Msn}{\Nn}.
\end{align}
Now, it holds that each of the $\Nn$ nodes with high effort interacts with a
node of low effort with probability $\pnmeets$ at rate $\tau$. Then, with
probability $\pns$ a node with high effort takes up the low effort. This causes the number of links between pairs of nodes with low effort to increase by the number of neighbors with low effort of the
formerly high-effort node,
\begin{align}
\frac{d\Ms}{dt}\sim\tau\Nn\pnmeets\pns\Kns.\label{eqn:imitation_increase}
\end{align}

A third term that governs the time evolution of $\Ms$
is given by its decrease 
caused by nodes with low effort that imitate the high effort. If a node $v_i$ with the low effort $E_i=\Es$ 
interacts with a node $v_j$ having the high effort $E_j=\En$ and $v_i$ then
imitates the effort of $v_j$, the total number of links connecting two
nodes with low effort decreases by the number of $v_i$'s neighbors $v_k$
that are showing the low effort $E_k=\Es$ as well. Following from an analogous
argument as given above, this number is given by $\kss(v_i)$. Again we assume the
number of neighbors $v_k$ with $E_k=\Es$ of a node $v_i$ with
$E_i=\Es$ to be approximated by its average,
\begin{align}
	\kss(v_j) = \Kss=\frac{2\Ms}{\Ns}.
\end{align}
With rate $\tau$ each of the $\Ns$ nodes with low effort
interacts with a node showing the high effort $\En$ with probability
$\psmeetn$. With probability
$\psn$ a node with low effort imitates the high effort which causes a
decrease in $\Ms$ by the average number of low-effort neighbors $\Kss$ of the node that is imitating
the high effort, 
\begin{align}
\frac{d\Ms}{dt} \sim -\tau
\Ns\psmeetn\psn\Kss.\label{eqn:imitation_decrease}
\end{align}
Putting together Eqs.~\eqref{eqn:adaptation},~\eqref{eqn:imitation_increase}
and~\eqref{eqn:imitation_decrease} gives the time evolution of the number of
links between nodes of low effort as
\begin{align}
\frac{d\Ms}{dt} &= \tau(\Nn\pnmeets\pns\Kns-\Ns\psmeetn\psn\Kss)\nonumber\\
&\quad+\rho\Ns\psmeetn.\label{eqn:dMs_dt}
\end{align}  
Plugging the definitions of $\Kss$ (Eq.~\eqref{eqn:Kss}) and $\Kns$
(Eq.~\eqref{eqn:Kns}) into Eqn.~\eqref{eqn:dMs_dt} and normalizing with the total number of nodes $N$ yields
the time evolution of the per node density of links between nodes of low
effort
\begin{align}
\frac{d\ms}{dt}=\tau(\pnmeets\pns\msn-2\psmeetn\psn\ms)+\rho\ns\psmeetn
\end{align}
which is again independent of $N$. Due to the symmetry of
the system, the time evolution of the per node density $\mn$ of links between nodes
with high effort then immediately follows as
\begin{align}
\frac{d\mn}{dt}=\tau(\psmeetn\psn\msn-2\pnmeets\pns\mn)+\rho\nn\pnmeets.
\end{align}

For the time evolution of the average stock of nodes with low
and high effort $\mus$ and $\mun$ we already found in Eqs.~\eqref{eqn:stock1}
and~\eqref{eqn:stock2} that 
\begin{align}
d\mus&=dt(\mus(1-\mus-\Es)-\mus^{(2)})+\deltans\\
d\mun&=dt(\mun(1-\mun-\En)-\mun^{(2)})+\deltasn.
\end{align}
The general forms of $\deltans$ and $\deltasn$ are (see
Eq.~\eqref{eqn:deltans} and~\eqref{eqn:deltasn})
\begin{align}
\deltans=\frac{\mun-\mus}{\ns}d\nns\\
\deltasn=\frac{\mus-\mun}{\nn}d\nsn.
\end{align}
For the case of an adaptive network, $d\nns$ ($d\nsn$) is given
by the first (second) term in Eq.~\eqref{eqn:dns_adaptive_general}:
\begin{align}
\deltans=\frac{\mun-\mus}{\ns}\tau\nn\pnmeets\pns\\
\deltasn=\frac{\mus-\mun}{\nn}\tau\ns\psmeetn\psn,
\end{align}
with the probabilities $\pnmeets$ and $\pnmeets$
(Eqs.~\eqref{eqn:adaptive_psmeetn} and~\eqref{eqn:adaptive_pnmeets}) as defined above and $\pns$ and
$\psn$ being the same as for the static model (Eqs.~\eqref{eqn:static_pns} 
and~\eqref{eqn:static_psn}).

To summarize, the set of five coupled differential equations that represent the
adaptive network model's macroscopic dynamics is given as
\begin{align}
\frac{d\ns}{dt} &= \tau (\nn\pnmeets\pns - \ns\psmeetn \psn) \label{adaptive_1}\\
\frac{d\ms}{dt} &= \tau (\pnmeets \pns \msn- 2 \psmeetn \psn \ms) + \rho \ns \psmeetn\\
\frac{d\mn}{dt} &= \tau (\psmeetn \psn \msn- 2 \pnmeets \pns \mn) + \rho \nn\pnmeets \label{adaptive_3}\\
\frac{d\mus}{dt} &= \mus(1 - \mus - \Es) + \tau\frac{\nn}{\ns}(\mun - \mus) \pnmeets \pns\\
\frac{d\mun}{dt} &= \mun(1 - \mun - \En) + \tau\frac{\ns}{\nn}(\mus - \mun) \psmeetn \psn. \label{adaptive_2}
\end{align}
It is important to note that in most previous works on adaptive networks a
closed set of macroscopic equations is obtained by assuming that links in the
network are drawn at random and interactions take place between nodes that are
connected by them~\cite{demirel_moment-closure_2014,Gross2006}. In this work
nodes, not links, are randomly drawn and initiate an interaction with its
neighboring nodes. This subtle difference changes the effective time scale of
the system. Specifically, in our model only a maximum of $N$ out of all $M$
links are affected by interactions between nodes during the same time as all
$M$ links would be considered if interactions take place by randomly drawing
links in the network. In other words, in our model it takes $M/N$ times longer
to achieve the same number of updates, as one would obtain by considering
per-link interactions. 

For this system, the stable fixed point's ${\ns}_0$-component can be obtained
numerically for different combinations of $\phi$ and $\DT$ (Fig.~\ref{fig:fig_03}
(B)). The results are again in good agreement with the numerical simulations
and imply that for every choice of $\DT>0$ there actually exists an appropriate
choice of $\phi \in [\pone, \ptwo]$ so that all nodes are likely to adopt the
effort level $\Es$.  The lower bound of the optimal rewiring probability
$\pone$ can be obtained by utilizing Eq.~\eqref{eqn:fp3} and the linear
relationship between $\pone$ and $\DT$ for a fixed rate of social updates
$\tau$ that lead to imitation as given in Eqn.~\eqref{eqn:tau_adaptive}. We
thus find $\pone(\DT,\DE) = \ptwo (1- \frac{2-2\DE^2}{1+\DE^2} \DT) \for 0<\DT
< \frac{1+\DE^2}{2-2\DE^2}$ and $\pone(\DT,\DE)=0$ otherwise. The upper bound
$\ptwo$ at which the network fragments is obtained from a numerical bifurcation
analysis as $\ptwo \approx 0.89$. The result is in good agreement with previous
findings on the fragmentation threshold in adaptive networks for similar
average degree $\overline k$~\cite{silk_exploring_2014,bohme_analytical_2011}.
We find, however, that the computed fragmentation threshold $\ptwo$ is larger
than what is expected from the numerical simulations (Fig.~\ref{fig:fig_03}
(A)). This can either be due to the fact that moment closure as well as
mean-field approximations are known to provide only rough estimates of the
fragmentation threshold~\cite{demirel_moment-closure_2014} or because finite
size effects in the numerical simulations cause the system to fragment for
smaller values of $\phi$ than it would be expected for the limiting case
$N\rightarrow \infty$ that is considered in the macroscopic approximations.  A
more detailed study of the network fragmentation and the corresponding
threshold $\ptwo$ is a subject of future research.

\subsection{Consistency between approximations}
\begin{figure}[t]
	\includegraphics[width=.9\linewidth]{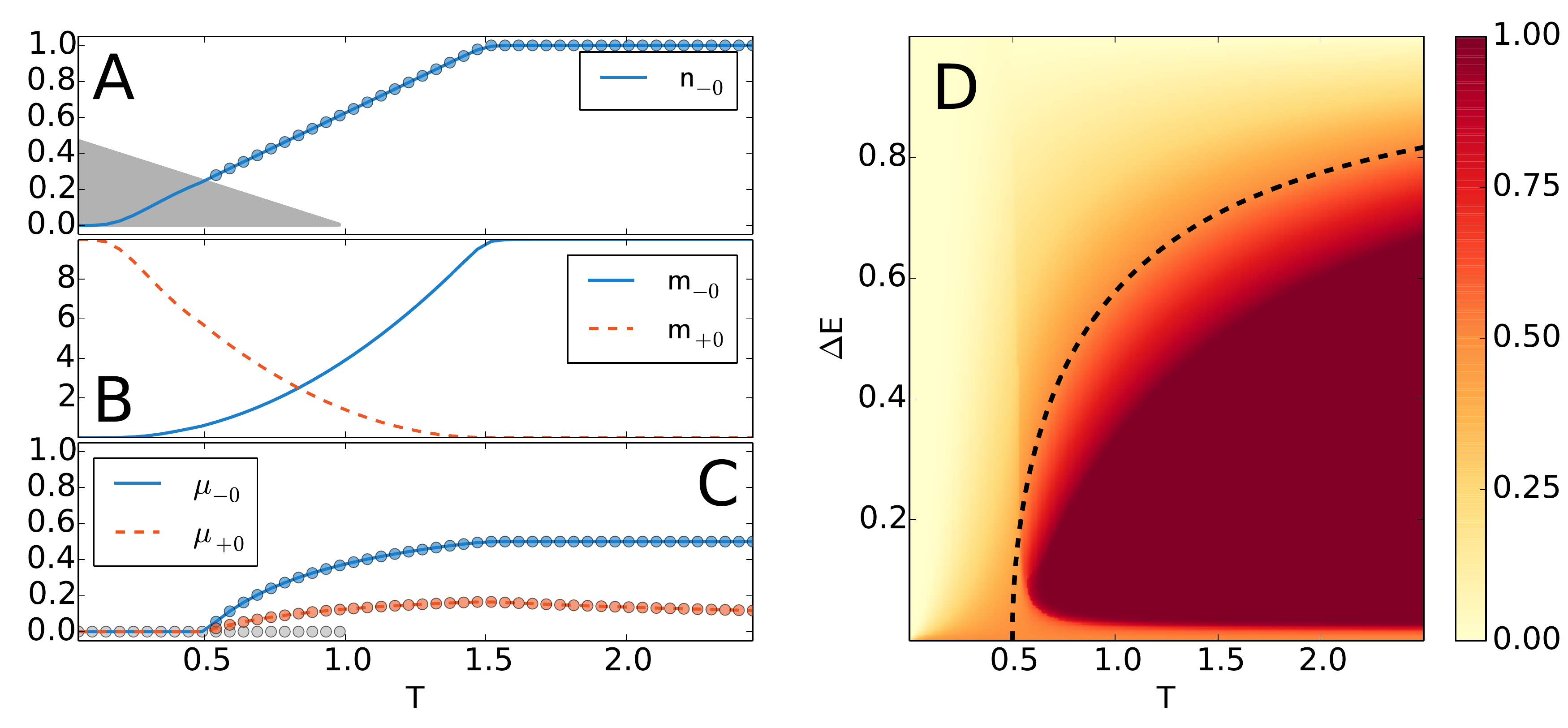}
	\caption{(Color online) (A)--(C) The dependence of the adaptive (solid lines) and static
		model's (transparent scatter) stable fixed point on the expected waiting
		time $\DT$ for fixed parameters $\phi=0$ and $\DE=0.5$.  (D) The adaptive
		model's stable fixed point's ${\ns}_0$-component indicating the fraction of
		nodes with effort $\Es$ in the consensus state as a function of the two
		parameters $\DT$ and $\DE$ again for $\phi=0$. The dashed line indicates
		the value of the critical waiting time $\T$ obtained from the set of
		differential
		equations~\eqref{eqn:05_dns_final}--\eqref{eqn:05_dmun_final}.}\label{fig:fig_05}
\end{figure}

To illustrate the consistency of the set of differential equations describing
the static setting~\eqref{eqn:05_dns_final}--\eqref{eqn:05_dmun_final} and the
adaptive case~\eqref{adaptive_1}--\eqref{adaptive_2}, we set $\phi=0$ in the
latter, compute its fixed points numerically and compare them with the static
setting's fixed points~\eqref{eqn:fp3} and~\eqref{eqn:fp4}.
Figure~\ref{fig:fig_05} (A-C) shows the different components
of the stable fixed points as a function of the control parameter $\DT$ for a
fixed $\DE=0.5$. The components ${\ns}_0$, ${\mun}_0$ and ${\mus}_0$ align
perfectly well for the static and the adaptive case. The gray shaded area in
Fig.~\ref{fig:fig_05} (A) indicates a center manifold for which the system's
stability cannot be assessed by standard linear stability analysis. However,
numerically integrating the set of differential equations yields the expected
behavior of $\ns(0)\rightarrow 0$ as $\DT\rightarrow 0$. 
Figure~\ref{fig:fig_05} (D) displays again the ${\ns}_0$-component of the adaptive
model's stable fixed point for $\phi=0$ and different combinations of $\DT$ and
$\DE$. The results match those of Fig.~\ref{fig:fig_01} (B). Hence, the system of
dynamic equations~\eqref{adaptive_1}--\eqref{adaptive_2} can be interpreted as a
consistent generalization of
Eqs.~\eqref{eqn:05_dns_final}--\eqref{eqn:05_dmun_final}.

\section{Conclusions}\label{sec:conclusion}
We have introduced a model to describe emerging structure formation from the
interplay of dynamics of and on networks manifested by the co-evolution of
social dynamics on the one hand and resource dynamics on the other hand. An
adaptive voter model has been coupled to a set of logistic growth models, such that
the state of the dynamic variables influences the imitation (i.e.\ social trait
adoption) processes in the underlying social network which take place according
to differences in harvest or payoff.  We have derived rate equations for the
system's macroscopic variables and demonstrated that the resulting system of
differential equations yields stable fixed points which are in good agreement
with the results from numerical simulations. 

Our paradigmatic example illustrates that the interplay between both types of
network dynamics gives rise to a variety of new phenomena, which have not been
observed so far when only studying either of the two aspects.  We have mainly
found that the rate of interactions in the network determines the expected
linear stability of the growth model's fixed points. However, for each choice
of interaction rate there exists an appropriate range of the adaptive rewiring
frequency so that the expected fraction of, e.g., nodes with effort $\Es$ can
be maximized.  Notably, the subset of differential
equations~\eqref{adaptive_1}--\eqref{adaptive_3} provides a general description
of imitation and adaptation dynamics on a social network with binary states of
nodes and symmetric imitation rules. Hence, it is applicable to study many
other problems as long as the imitation probabilities $\psn$ and $\pns$, which
do not have to be constant for all times, are chosen appropriately. 

The proposed model also raises questions that need to be addressed in future
research. In the course of the macroscopic approximation we have assumed all
moments of higher order in stocks and network structure to vanish such that the
set of differential equations could be closed. The results have been shown to
be in good agreement with numerical simulations. However, a more in-depth
analysis of whether the inclusion of higher order moments would enable us to
reproduce the steep transition between the two regimes of predominance of low-
or high-effort nodes remains a relevant research questions. 
We also aim to estimate more thoroughly the
critical waiting time $\T$ at which the observed phase transition takes place
and therefore investigate the expected time at which the low effort provides
more harvest than the high effort given that no interaction between the nodes
took place so far. Finally, we aim to obtain data from agricultural studies on,
e.g., water usage or harvest exploitation of resources, to test the findings
and insights that we have obtained from our co-evolutionary model with respect
to real-world phenomena.

\begin{acknowledgments}
This work was carried out within the framework of PIK's COPAN project.  MW has
been supported by the German Federal Ministry for Science and Education via the
BMBF Young Investigators Group CoSy-CC$^2$ (grant no. 01LN1306A). JFD and WL
acknowledge funding from the Stordalen Foundation and BMBF (project GLUES) and
JK acknowledges the IRTG 1740 funded by DFG and FAPESP\@. We thank R.V. Donner
for helpful comments and suggestions on the manuscript and R. Grzondziel
and C. Linstead for help with the IBM iDataPlex Cluster at the Potsdam
Institute for Climate Impact Research.
\end{acknowledgments}

\end{document}